\documentclass[useAMS]{mn2e}
\usepackage{psfig}

\newif\ifAMStwofonts

\def\be{\begin{equation}}
\def\ee{\end{equation}}

\def\msun{{M_\odot}}

\def\etal{{\it et al.~}}

\def\gtsima{$\; \buildrel > \over \sim \;$}
\def\ltsima{$\; \buildrel < \over \sim \;$}
\def\prosima{$\; \buildrel \propto \over \sim \;$}
\def\gsim{\lower.5ex\hbox{\gtsima}}
\def\lsim{\lower.5ex\hbox{\ltsima}}
\def\simgt{\lower.5ex\hbox{\gtsima}}
\def\simlt{\lower.5ex\hbox{\ltsima}}

\def\simpr{\lower.5ex\hbox{\prosima}}

\def\etal{{\frenchspacing\it et al. }}
\def\ie{{\frenchspacing\it i.e. }}

\def\be{\begin{eqnarray}}
\def\ee{\end{eqnarray}}

\def\CR{{\tt CRASH}~}

\title[The Proximity Effect Around High Redshift Galaxies]{The Proximity Effect Around High Redshift Galaxies}

\author[A. Maselli, A. Ferrara, M. Bruscoli, S. Marri \& R. Schneider]{
A. Maselli$^{1}$
,
A. Ferrara$^{2}$
,
M.Bruscoli$^{1}$
,
S. Marri$^{2}$
and
R. Schneider$^{3}$\\
$^1$ Dipartimento di Astronomia, Universit\'a di Firenze, Largo
Enrico Fermi 5, 50125 Firenze, Italy\\
$^2$ SISSA/International School for Advanced Studies, via Beirut 2-4,
34014 Trieste, Italy\\
$^3$ Osservatorio Astrofisico di Arcetri, Largo
Enrico Fermi 5, 50125 Firenze, Italy
}

\date{October 2003}

\pagerange{\pageref{firstpage}--\pageref{lastpage}}
\pubyear{2003}
\begin{document}

\maketitle
\label{firstpage}

\begin{abstract}
Recent observations have shown that the intergalactic medium (IGM)  
is more transparent to Ly$\alpha$ photons close to Lyman Break Galaxies (LBGs) than 
at large distance from them, \ie a proximity effect. Cosmological simulations 
including winds from LBGs have been so far unable to explain this trend.     
By coupling such simulations with the radiative transfer code {\tt CRASH}, we 
investigate whether the addition of the ionizing radiation emitted by LBGs 
can increase the 
transmissivity by decreasing the neutral hydrogen fraction in the inner Mpc of
the galaxy halo.  The transmissivity as a function of distance is roughly
reproduced only if LBGs are identified with dwarf galaxies (with masses
$\simlt 10^9 M_\odot$) which are undergoing a vigorous ($50 M_\odot$~yr$^{-1}$)
burst of star formation. Similar star formation rates in larger galaxies 
are not sufficient to overwhelm the large recombination rates associated with
their denser environment. If so, photoionization partly reconciles
theory with observations, although we discuss a number of uncertainties affecting
both approaches. 

\end{abstract}


\section{Introduction}
In a recent study, Adelberger {\it et al} (2003, hereafter A03) have analyzed 
a sample composed of 8 high-resolution QSOs spectra at $3.1<z<4.1$ and 
spectroscopic redshifts for 431 Lyman-break galaxies (LBGs) at slightly lower 
$z$, to study the intergalactic medium (IGM) close to high-$z$ galaxies and 
detect imprints of galaxy-IGM interplay.    
Comparing the location of galaxies to the absorption lines in QSOs spectra, 
the authors found that the mean Ly$\alpha$ transmitted flux generally 
increases with the line-of-sight impact parameter between the foreground galaxies 
and the background QSOs ($\Delta r$), reaching the mean transmissivity of 0.67 
at $\Delta r \gsim 6$ $h^{-1}$ comoving Mpc; hovever, the trend is reversed
within about $\sim$ $1 h^{-1}$ comoving Mpc of the galaxy. 
This result indicates that the IGM contains less neutral hydrogen close 
to LBGs, and hence is more transparent to Ly$\alpha$ photons, than at large 
distance from them.  \\
Similar observations at lower redshift show an opposite behavior. Lanzetta 
{\it et al} (1995), Chen {\it et al} (1998) and Pascarelle {\it et al} (2001) 
find that galaxies, at $z < 1$, with an impact parameter $\Delta r 
\lsim 180 h^{-1}$ comoving kpc are much more often associated with Ly$\alpha$ 
absorption lines than those at larger impact parameters. Furthermore the 
result of VLT/UVES study of the Ly$\alpha$ forest in the vicinity of the 
LBG MS1521-cB58 at $z=2.724$ (Savaglio, Panagia \& Padovani 2002) shows 
an absorption {\it excess} close to the galaxy.

A03 suggest that the relative lack of neutral hydrogen at small distances 
from LBGs is possibly due to the dynamical feedback from supernova-driven 
winds, which may be able to accelerate the surrounding gas up to 
velocities exceeding the escape velocity of the galaxy potential well, 
driving plausibly the intergalactic 
material out to a distance $> 0.5 h^{-1}$ comoving Mpc. 

This scenario has been investigated with numerical simulations 
by different groups: Croft {\it et al} (2002), Bruscoli {\it et al} (2003),
and Kollmeier {\it et al} (2003). 
Despite the completely independent prescriptions adopted to model 
supernova-driven winds in these studies, none of them
were able to reproduce A03 results on the entire distance range. 
Unless feedback schemes are still far too crude in SPH simulations, 
the discrepancy indicates that the energy injection from supernova 
explosions is not able to displace the IGM around galaxies 
to distances $\ge 0.5 h^{-1}$ comoving Mpc. The most likely explanation
is that the ram-pressure of the accreting/infalling gas slows down and 
effectively confines winds to relatively modest distances from the explosion 
sites (Bruscoli \etal 2003).

Local photoionization by galaxies themselves provide the only known 
alternative mechanism able to decrease the neutral hydrogen fraction in 
the LBG environment. 
The first attempts to account for local photo-ionization in numerical 
simulations have been performed 
using simplified analytical techniques and/or ad-hoc prescriptions (A03, 
Kollmeier \etal 2002, Croft \etal 2002); these studies have lead to  
conclude that this process cannot account for the observed IGM transparency. 
However, fully self-consistent and detailed radiative transfer 
calculations are required before a final conclusion is drawn. 
  
In this Letter we investigate the role of local photoionization 
using high resolution numerical simulations performed with the radiative 
transfer code \CR; the full description of the code and of the tests 
made are given in Maselli, Ferrara \& Ciardi (2003).

\section{Simulations}
We have performed radiative transfer simulations in order to evaluate 
self-consistently the proximity effect due to the ionizing 
radiation emitted by high redshift galaxies on the surrounding IGM. 
We use the numerical code \CR (Maselli \etal 2003) which allows to
calculate the effects of ionizing radiation on a precomputed cosmological 
density field, specifically obtained from SPH simulations fully described
elsewhere (Marri \etal 2003) and already used by Bruscoli \etal (2003). 
In the following we give the details of the two simulations. 

\subsection{Hydrodynamics}
The simulation has been carried out with an improved version 
of the SPH code GADGET2 (Springel \etal 2001), designed for the detailed 
study of galaxy formation (Marri \& White 2002). 
Such multi-phase (MSPH) scheme is based on an explicit separation of 
protogalactic 
gas into diffuse and dense (star forming) components and 
includes new prescriptions for a realistic treatment of stellar feedbacks.
With respect to the standard GADGET2 version, this new implementation 
is able to suppress star formation, to reheat cold gas and to drive outflows 
from galactic disks.  The presence of a UV 
background (UVB) produced by QSOs and filtered through the IGM is also 
included.
The shape and amplitude of the UVB are taken from Haardt \& 
Madau (1996) and its impact on the IGM temperature  and ionization structure  
is calculated as in Katz \etal (1996). 

The simulation uses  $128^3$ particles in a 10.5$h^{-1}$ comoving Mpc box  
and assumes a $\Lambda$CDM 
cosmological model with $\Omega_0=0.3$, $\Omega_\Lambda=0.7$, 
$\Omega_b=0.04$ and $h=0.7$; the initial density perturbations 
power spectrum is cluster-normalized ($\sigma_8=0.9$) and periodic 
boundary conditions are assumed. 
Galaxies are identified by applying the HOP group finding
algorithm (Eisenstein \& Hut 1998) to the baryonic
components of the simulation. Each galaxy is assigned   a 
star formation rate (SFR) calculated by considering the properties of 
all particles within the group. For each of them 
we first check if the conditions under which a 
gas particle can form stars are satisfied and 
we calculate the rate at which its gas is converted into stars. 
For more details we defer the reader to Marri \& White (2002).

For consistency sake, we use the same output as in Bruscoli \etal (2003).
There we showed that, although on large scales the simulation reproduces 
a number of statistical properties of the IGM, it does not match the 
transmitted Ly$\alpha$ flux inside 1 $h^{-1}$ comoving Mpc as measured 
by A03.  Such an output has now been post-processed using \CR 
to self-consistently include the effects of local 
photoionization produced by the galaxies identified in the 
computational volume.

\subsection{Radiative Transfer}

\begin{figure*}
\centerline{\psfig{figure=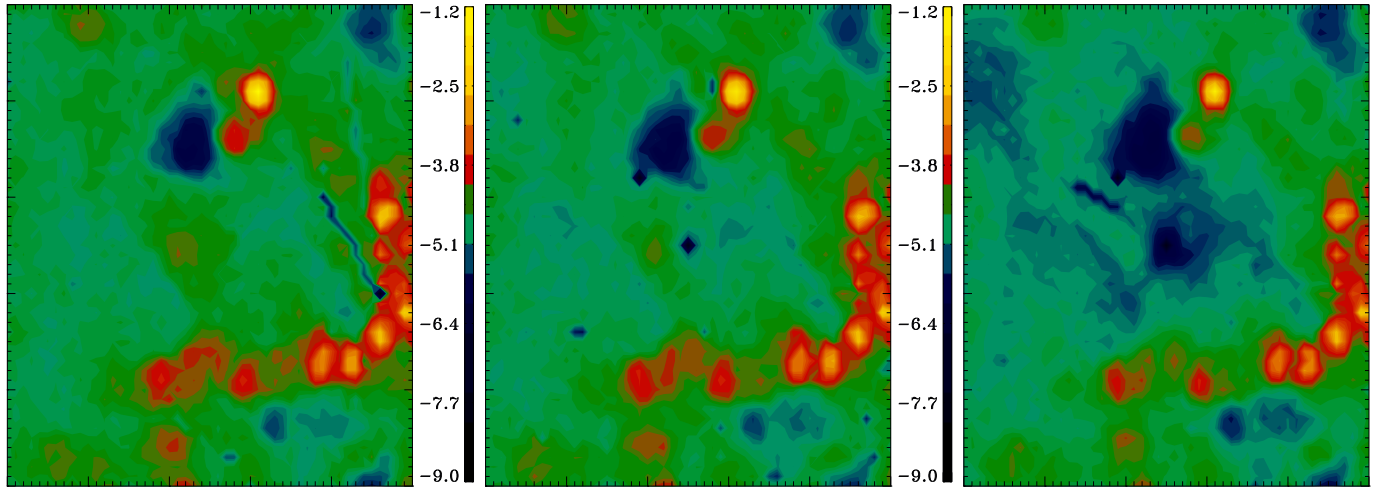,height=7.cm}}
\caption{Neutral hydrogen fraction maps in a slice of about 4$h^{-1}$ 
comoving Mpc by side, centered on the most massive and most 
luminous galaxy in the computational volume. The three slices 
represent, in a logarithmic scale, the final outputs of run A 
(left panel), run B with the galaxy SFR derived from the MSPH simulation 
(central panel), and  run B with the SFR boosted by a factor of 10 
(right panel).}
\label{fig01}
\end{figure*}

The radiative tranfer (RT) simulation has been performed on 
the outputs at $z=3.27$ of the MSPH simulation above, 
to allow a direct comparison to the observational data available 
around this redshift (A03).  

The RT is simulated running the 3-D RT code \CR, whose 
description is fully reported in Maselli \etal (2003). 
The code works on precomputed H/He density fields assigned in a 3-D 
cartesian grid, and calculates the time evolution 
of the temperature and of the ionization structure. 

To include the effect of local photoionization on the 
cosmological configuration obtained from the MSPH simulation, it is necessary to 
map the particle data on a 3-D cartesian grid. We choose a grid composed of 
128$^3$ cells, which corresponds to a spatial resolution of 
$\Delta x \sim 82$ kpc $h^{-1}$ comoving. 
At each grid point we assign a value for the mass density ($\rho$), 
temperature ($T$), neutral hydrogen fraction 
($x_{\rm H^0}=n_{\rm H^0}/n_{\rm H}$) and free electrons fraction 
($x_e=n_{\rm e}/(n_{\rm H}+n_{\rm He})$), performing a standard SPH 
smoothing using the 32 neighbour SPH particles closer to the grid point. 
We next locate the 398 galaxies identified 
inside the simulation box on the cartesian grid, converting the 
lagrangian spatial coordinates into cartesian coordinates. \\
The emission properties of the galaxies have been derived using 
the Starburst99 code (Leitherer \etal 1999), a web based 
software and data package designed to model spectrophotometric and related 
properties of star-forming galaxies. We have assumed a continuous star 
formation law and a Salpeter IMF with $\alpha=2.35$, normalized in a  mass 
range [1,100]$\msun$; we also take Z$_\odot$/5 as a characteristic 
metallicity for all the sample. 
These values are suggested by the analysis 
of the LBGs UV continuum and the H$_\alpha$ and H$_\beta$ emission lines by 
Pettini \etal (2001). As the physical connection of LBGs with galaxy type/mass 
is unclear  (Primack \etal 2003), we assume for all galaxies in the box 
similar scalings. Running Starburst99 under the above conditions, 
we obtain the galaxy SED corresponding to the derived SFR of each galaxy.  
We finally assume a value for the ionizing photon escape fraction of
50\%. 
This value represents the upper limit of the allowed range (see discussion in 
Ciardi, Bianchi \& Ferrara 2002) from a set of observational data. 

The computed grid data are assigned as initial conditions for the \CR run. 
The number densities of hydrogen and helium nuclei are calculated 
assuming mass fractional abundances  $f_H$=0.76 and $f_{He}$=0.24, 
respectively. Temperature and hydrogen ionization fractions are 
initialized as in the MSPH simulation output; for helium
we assume ionization fractions  
$x_{\rm He^+}=x_{\rm H^+}$ and
$x_{\rm He^{++}}=(x_e - x_{\rm He^+}-x_{\rm H^+})/2$.
The physical duration of a simulation run has been set to 
$t_s=6 \times 10^6$ yrs, \ie several times 
the estimated box-crossing-time for the ionization front produced by 
a typical source in the volume.

Nearby galaxies the mean gas temperature is above 10$^5$ K, 
due to the energy injection into the IGM produced by outflows.  
The cooling time of such gas in the simulation is longer 
than $t_s$, with a minimum value equal to 1.5$\times$10$^8$~yrs. 
As the temperature does not change significantly 
during the RT simulations, its evolution has not been solved for  
to reduce the computational time.

Two different RT simulations have been run.
In the first one only the UVB radiation is included (run A), 
while in the second one, the radiation from all galaxies in the 
simulation volume is added to the UVB (run B). 
In this way it is possible to compare consistenly the two simulations 
to infer the role of galaxies and distinguish the contribution to 
the proximity effect due to local photoionization. 
As considerable uncertainty is present in the determination of SFRs
of LBGs, we also studied a case similar to run B in which however SFRs are
boosted {\it ad hoc} to match the observed values.

The total ionizing energy, injected in the computational 
volume, is discretized in $N_p=2 \times 10^8$ photon packets for 
run A and in $N_p = 6 \times 10^8$ packets for run B:  
roughly, this corresponds to $10^6$ photon packets per galaxy 
and $2 \times 10^8$ photon packets for the UVB.

\section{Results}

As an example, Fig.1 shows a comparison among the final outputs 
of run A (left panel) and run B (central and right panels). 
The slices represent the logarithm of the neutral hydrogen fraction on a slice  
of a zoomed region of about 4$h^{-1}$ comoving Mpc by side, centered 
on the most massive galaxy in the computational volume. 
The two maps from run B have been derived by assigning to such 
a galaxy different SFR: in the central panel the SFR is about 
30 M$_\odot$ yr$^{-1}$ as derived from the MSPH simulation, while 
in the right panel it has been boosted by a factor of 10.   
The effect of the local photoionization produced by 
the galaxy is clearly visible in both cases and depends strongly 
on its luminosity, as expected. In the central panel the imprint of the 
galaxy on the surrounding gas is feeble and the content of neutral hydrogen 
in a surrounding region of 0.5 h$^{-1}$ comoving 
Mpc by side, is reduced on average by factor of 2. The luminosity 
associated with a SFR of 30 M$_\odot$ yr$^{-1}$ 
is too low to affect significantly the high density environment of such 
a massive galaxy (M =$1.5\times 10^{11}$M$_\odot h^{-1}$).
When the SFR is boosted by a factor of 10 (right panel), 
the region affected by the galaxy  UV flux extends up to a distance of about 
1$h^{-1}$ comoving Mpc. In the inner region the neutral hydrogen fraction 
is depleted at least by an order of magnitude. In the outer regions the 
effect of the local photoionization is progressively reduced and 
vanishes at distances $\ge 1 h^{-1}$ Mpc, where the UVB dominates the local 
UV flux.
The decrease in the neutral hydrogen fraction visible in other regions
far from the center is produced by other galaxies, included 
in run B as ionizing sources. 

To estimate the radius of the ``sphere of 
influence'' produced by the local emission from a typical galaxy we 
study the distribution of the ratio between the UV 
flux produced by the galaxies and the UVB flux, ${\cal R}=F_{gal}/F_{bkg}$, 
in the box. 
Fig.2 shows the histogram derived from the final output of run B. The 
distribution has a mean value $\overline{\cal R}=2.64 
\times 10^{-2}$ and a median value ${\cal R}_M=1.83 \times 10^{-2}$. 
These values are to be considered as lower limits to the contribution 
of galaxies to the UVB, as larger boxes would be required to properly 
estimate such quantity.
The volume in which ${\cal R}>1$, \ie where the local emission by galaxies 
dominates the foreground UVB, 
is about $4.8 \%$ of the total computational volume. 
This fraction corresponds to a mean radius of influence for a typical 
galaxy $R_i$, given by: 
\be
\label{eq1}
R_i=\left(\frac{4}{3\pi}\frac{N_c({\cal R}>1)\times V_c}{N_g}\right)^{1/3} 
\simeq 0.4 h^{-1} {\rm Mpc}
\ee
where $N_c$ is the number of cells where ${\cal R}>1$, 
$V_c$ is the volume of a single cell and $N_g=398$ is the number of 
galaxies in the simulation box. 

To allow  a direct comparison of the simulation results 
with the observational data we construct synthetic Ly$\alpha$ spectra as 
described in Bruscoli \etal, by tracing random LOS through the box. 
We then estimate the mean Ly$\alpha$ transmitted 
flux as a function of the impact parameter $\Delta r$, computing the 
average on all the pixels at a distance from a galaxy in a given interval 
centered on $\Delta r$. The results are plotted in Fig.3 and 
compared with the data, represented by the black points.

\begin{figure}
\centerline{\psfig{figure=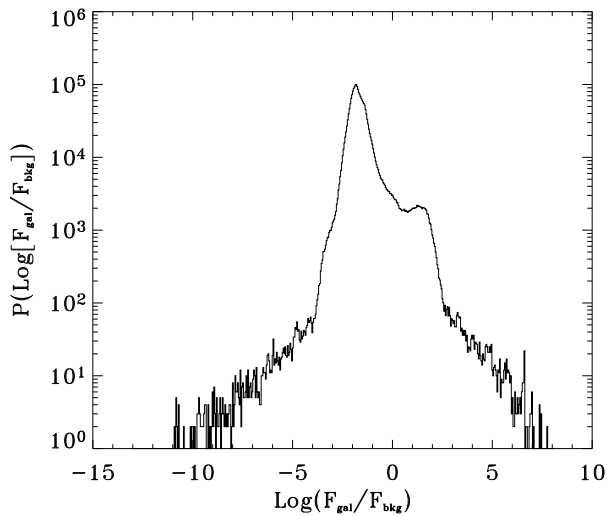,height=7.5cm}}
\caption{Distribution of the ratio between the UV 
flux produced by the galaxies and the UVB flux, ${\cal R}=F_{gal}/F_{bkg}$, 
in the box, as derived from the final output of run B.} 
\label{fig2}
\end{figure}

\begin{figure*}
\centerline{\psfig{figure=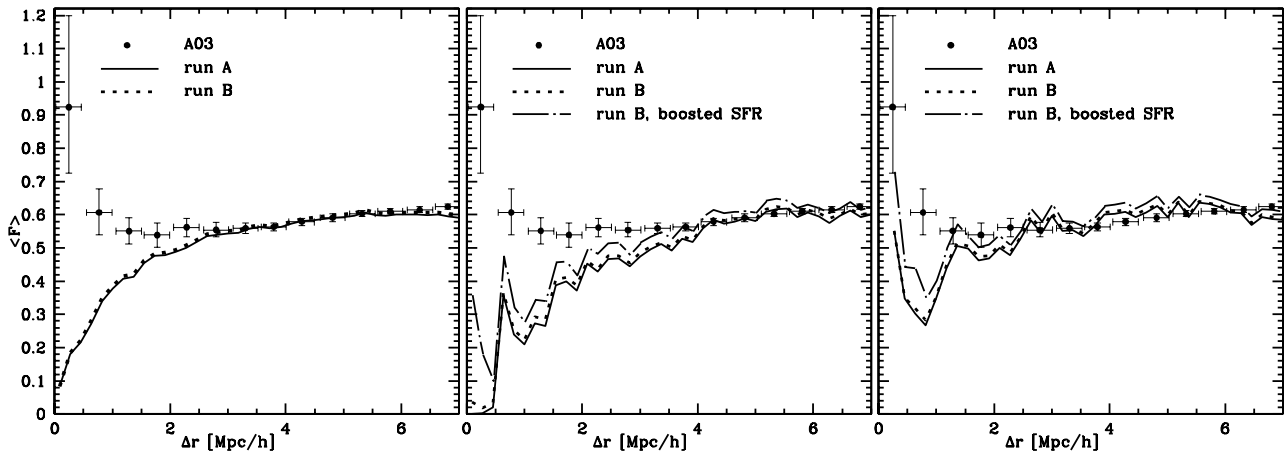,height=6.cm}}
\caption{Mean Ly$\alpha$ flux at different impact parameters $\Delta r$, 
computed as the average values among all the pixels at a distance from a 
galaxy in a given interval centered on $\Delta r$. The curves in the three 
panels are derived using different samples 
of galaxies: all the 398 galaxies (left panel), galaxies with mass above 
$2 \times 10^{10}$ M$_\odot$ (central panel) and those  with mass below 
$9.3 \times 10^8$ M$_\odot$ (right panel).}
\label{fig3}
\end{figure*}

The three panels show the mean Ly$\alpha$ flux computed on different samples 
of galaxies: all the 398 galaxies (left panel), galaxies with mass above 
$2 \times 10^{10}$ M$_\odot$ (central panel) and those  with mass below 
$9.3 \times 10^8$ M$_\odot$ (right panel).

Solid, dotted and dotted-dashed lines are derived from the 
outputs of run A, run B and from a simulation 
analogous to run B, but with SFRs boosted {\it ad hoc}. 
The difference between run A and run B in the left panel is marginal:
we conclude that the mean effect of the local photoionization on the 
Ly$\alpha$ transmitted flux is negligible when all galaxies are included 
in the analysis. 

Massive galaxies are the best candidates for LBGs due to their high 
luminosity and clustering properties.  
The SFR derived for the most massive galaxies in the MSPH simulation are in 
the range $10\div30$ M$_{\odot} {\rm yr}^{-1}$, the highest values in 
the simulation. 
Due to their high luminosity, one would expect them to produce a 
strong impact on the ionization of the surrounding gas. However 
this is not the case. Looking at the solid line in the central panel, 
obtained by neglecting the 
local emission, we note that the high density characterizing the 
environment of massive galaxies suppresses the mean transmitted flux with 
respect to the mean trend obtained for all the galaxies in the simulation
(solid line in the left panel). The UV radiation 
emitted by these galaxies, using the SFR derived from the MSPH simulation 
(dotted line in central panel), is not strong enough to enhance significantly 
the transparency of the surrounding IGM, because of the high recombination 
rate in the denser environment.
Higher values of SFR are necessary to produce a significant increase of the 
transmitted flux nearby massive galaxies. This can be seen from the boosted 
case, \ie with SFR in the range $100 \div 300$ M$_\odot$ yr$^{-1}$. Although on the high side,
the values are still plausible according to observations (Shapley \etal, 2001).
Despite the sharp increase in the mean Ly$\alpha$ transmitted flux, 
yet this simulation does not match the A03 data.
This result suggests that the mean transmissivity strongly depends on 
the galactic enviroment. 
The right panel shows the trend for the Ly$\alpha$ transmitted flux 
obtained selecting galaxies with mass $ \le 9.3 \times 10^8$ M$_\odot$. 
In this case the transmissivity of the gas at $\Delta r < 0.7 h^{-1}$ Mpc has 
an opposite trend with respect to the mean (left panel), and seems 
to follow the one observed by A03. The increasing flux at smaller 
distance from the galaxy reflects the lower density in these regions. 
Here the UVB flux can counteract the recombinations which occur on longer 
time scales. Nevertheless the SFRs associated with these galaxies are 
too small (of order $0.1 M_\odot$ yr$^{-1}$) to further increase the 
transmitted flux. Boosting the SFR up to values around 50 M$_\odot$ yr$^{-1}$ 
(dashed-dotted line), the increase 
in $\langle F \rangle$ is significant and closely approches the observed value 
in the innermost region. By no means though the flux around 1 Mpc can 
be matched. 

\section{Conclusions}
A recent study on the statistics of the Ly$\alpha$ forest in the vicinity 
of foreground galaxies has revealed an unexpected relative lack of neutral 
hydrogen in the inner Mpc $h^{-1}$ comoving from LBGs at $z \sim 3$. 
High energy supernova-driven winds able to displace the surrounding gas 
out to distances greater than $0.5 h^{-1}$ have been proposed as a viable explanation. 
However, several numerical studies failed in reproducing 
the observed effect of such superwinds via (M)SPH cosmological simulations. 
According to the simulations, the velocity of the gas closest to 
galaxies varies around  a characterisitc value of 40 km s$^{-1}$ for 
small objects, and around 150 km s$^{-1}$ for the biggest ones. These values 
seem to underestimate the high velocity gas, at 600 km s$^{-1}$, 
measured by A03 at a scale of a few kpc from the galaxy center. 
Nevertheless, the possibility for the wind to break through the galactic halo, and into the IGM 
up to distances greater than $1 h^{-1}$ comoving Mpc (even at such high velocity) 
seems quite unphysical if one takes into account the kinetic energy 
used to counteract the pressure of the accretion flow. 

Motivated by these reasons, we have studied the role of 
local photoionization in determining the Ly$\alpha$ absorbers statistics 
in the vicinity of galaxies, 
via numerical radiative transfer simulations. We post-processed an output 
at $z=3.27$ of a multiphase SPH simulation, in which a consistent treatment 
of supernova feedback is implemented, running the code \CR to account for 
the impact of the UV flux emitted by galaxies on the simulated IGM.  
We have derived the synthetic mean Ly$\alpha$ transmitted flux, as a function 
of the distance from a typical galaxy, for different samples of galaxies. 
We found that on average local photoionization has a negligible impact,  
\ie when all the galaxies present in the box are considered in the flux distribution
derivation. 

Local phoionization can affect significantly the 
transparency of the surrounding IGM only for particular conditions of the 
galactic environment. We find that in order to reproduce the trend inferred 
by A03, the gas around LBGs should be less dense than that surrounding 
massive galaxies seen in the simulation. The transmissivity as a function of distance is roughly
reproduced only if LBGs are identified with dwarf galaxies (with masses
$\simlt 10^9 M_\odot$) which are undergoing a vigorous ($50 M_\odot$~yr$^{-1}$)
burst of star formation. Similar and higher star formation rates in larger 
galaxies are not sufficient to overwhelm the large recombination rates 
that are occurring in their denser environment. 
This conclusion is somewhat at odd with the common wisdom that LBGs are 
predominantly massive systems, although different and equally valid arguments 
tend to identify them with the bursting population of dwarfs that we find 
necessary to explain the observed gas transparency.  
LBGs can be interpreted as outbursting dwarf galaxies (Bullock \etal 1999) 
if their abundance is due to an increased collision rate at high redshift;
although many of the simulated collisions have relatively small masses 
($\simlt  10^{10} M_\odot$), they tend to cluster about large-mass halos. 
They therefore exhibit strong clustering, similar to that observed.

The solve the puzzle, more data are needed. 
The poor statistics of A03 sample at small impact parameters, errors in measured redshifts and galaxy positions 
make the derived transmitted flux quite uncertain.

\label{lastpage}
\end{document}